\documentclass[12pt]{iopart}

\usepackage{cite}
\usepackage{iopams}
\usepackage{amsfonts}
\usepackage{amssymb}
\usepackage{bm}
\usepackage{graphicx}
\usepackage{datetime}
\usepackage{color}
\newcommand{\text}{\mathrm}
\usepackage{hyperref}
\hypersetup{colorlinks}
\definecolor{darkblue}{rgb}{0,0,1.0}
\hypersetup{ colorlinks,urlcolor=darkblue,linkcolor=darkblue,citecolor=darkblue}

\begin{document}
\newcommand{\figwidth}{0.95\columnwidth}
\newcommand{\ffigwidth}{0.4\columnwidth}
\newcommand{\unicamp}{Instituto de Fisica Gleb Wataghin, Universidade Estadual
de Campinas,
Rua S\'{e}rgio Buarque de Holanda, 777 Cidade Universit\'{a}ria 13083-859
Campinas, SP Brazil}
\newcommand{\unicamplimeira}{Faculdade de Ci\^{e}ncias Aplicadas, Universidade
Estadual de Campinas, 13484-350 Limeira, SP Brazil}

\title[Electronic localization at mesoscopic length scales]{Electronic
localization at mesoscopic length scales: 
different definitions of localization and contact effects in a heuristic DNA model}

\author{C.J. P\'{a}ez$^1$ and P.A. Schulz$^{1,2}$}

\address{$^{1}$\unicamp}
\address{$^{2}$\unicamplimeira}

\ead{\mailto{cjpaezg@ifi.unicamp.br}, \mailto{pschulz@ifi.unicamp.br}}

\date{1.74, compiled \today, \currenttime}

\begin{abstract}

In this work we investigate the electronic transport along model DNA molecules
using an effective tight-binding approach that includes the backbone on site
energies. The localization length and participation number are examined as a function of system size, energy dependence, and the
contact coupling between the leads and the DNA molecule. On one hand, the
transition from an diffusive regime  to a localized regime for short systems is
identified, suggesting the necessity of a further length scale revealing the
system borders sensibility.  On the other hand, we show that the lenght localization and participation number, do not depended of system size and contact
coupling in the thermodynamic limit. Finally we discuss possible length
dependent origins for the large discrepancies among experimental results for the
electronic transport in DNA sample.

\end{abstract}



\submitto{\EJP}

\section{Introduction}
\label{sec:intro}
The localization of all electronic states in disordered low dimensional systems
became a paradigm with the scaling theory of localization over thirty years ago \cite{AbrALR79}.
Nevertheless, this general picture of electronic localization
has been challenged by mainly two further developments. On one side, it has been
pointed out that correlations in disorder can lead to delocalization. Proposed
initially for specific short range correlation \cite{DunWP90}, it has been shown
that a variety of correlations, including long order ones \cite{MouCL04},
promotes
electronic delocalization. On the other hand, the continuous shrinking of solid
state devices turned the system size an important length scale. Hence,
electronic states normally localized in the thermodynamic limit, encounter a
situation in which the localization length (LL), is of the order of, or even
larger than, the system length \cite{GuhMW98}. More recently, the
possibility of macromolecular systems as a nanoelectronic framework, brought the
attention to the electronic properties of DNA strands under a wide range of
conditions. The
initial results of the search for unraveling the electronic transport properties
of DNA \cite{EndCS04,beratan97} caused surprise due to the variety of
findings, since DNA appeared to be either an insulator \cite{BockrathMSTGKWS02,BraESB98,PabMCH00}, a semiconductor
\cite{PorBVD00,CunCPD02,YooHLP01,feng02} or a metal \cite{fink99,rap01}. These initially controversial results suggested
eventually a complex scenario not yet completely understood.  The transport
properties of DNA may be affected by a quite long list of effects, either environmental
like interaction with substrates, sample dryness, counter ions effects; or
intrinsic like nucleotide sequencing \cite{WalJS10,GuoGHB08,gpc05,ShiRR08}.
In order to cope this variety of experimental results, numerous theoretical
works of electrical transport through DNA has been extensively studied, ranging
from molecular dynamics investigations, through \textit{ab initio} approaches \cite{PerRC10,MLC08},
heuristic tight-binding models \cite{MehA05,BagK07,Mac07}, as well as hybrid
methods \cite{GutCW09}, which attempt to bridge the characteristic length scales to the other
methods. 

These investigation efforts, on its turn, revealed an interesting
research stage, where different kinds of disorder, with and without
correlations, in systems of variable size in the mesoscopic scale are investigated by means of the
same heuristic models once used in numerical studies of electronic localization
in the thermodynamic limit. To mention but one of the findings in this scenario,
anomalous large LL are encountered for electronic states in
double strand (ladder) models in the presence of antisymmetric correlations
among on-site energies \cite{CarLM11}. 

This scenario of different kinds of disorder in designed structures at
mesoscopic dimensions may challenge the very definition of LL.
Indeed, the different definitions of electronic state localization found in the
literature, defined in the thermodynamic limit, may show distinct behaviors at mesoscopic system sizes, here an actual
relevant length scale. The aim of the present work is to generalize previous
findings which suggest an actual wave function extension over an order of
magnitude longer than the LL \cite{CarP08}. By comparing the
definitions of LL, participation ratio and the sensitivity to
boundary conditions \cite{KraM93}, an unambiguous estimate of a physical
thermodynamic limit
length can be defined for a rather general tight binding ladder model \cite{CunMRR07}.

\section{Theoretical formulation}
\label{sec:theory}
In the present work we address one of the various aspects in the long route
starting at the investigation of DNA electronic properties, their relations to the
transport properties and eventually ending at the proposal of nanoelectronic
devices: here we focus on the role of Lead/DNA/Lead coupling strength,
as an important parameter for studying the electronic transport through the DNA as function of systems
size.

In what follows we describe the general model chosen for this purpose, namely a
heuristic model within a tight-binding framework
including effective sites for the 
nitrogenated bases, as well as for sugar and phosphate backbone units, showing
relevant second nearest neighbor 
hopping among nitogenated bases \cite{ShiWHC12,CarpNR12}. Such a model captures general
physical features of the system under investigation and permits an unambiguous
addressing of the localization properties, from a fundamental point of view we
are interested in.

\begin{figure}[t]
\begin{center}
\includegraphics[width=0.5\columnwidth]{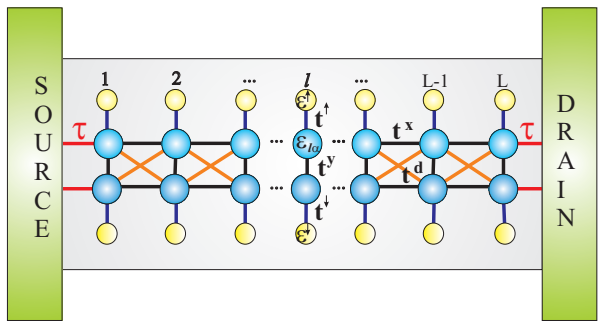}
\end{center}
\caption{Schematic representation of the DNA ladder model
attached to two semi-infinite electrodes at the left (source) and right (drain).
The nucleotide base pairs are depicted as blue circles representing the four
possible effective nucleotides: A, T, C, and G. Sugar-phosphate (backbone) effective sites are
depicted as yellow circles and electronic hoppings are shown as lines. Throughout
this work we will consider the limiting case of a completely random sequencing
along the central chains, maintaining solely the inter chain base pairing
(A-T and C-G).}
\label{fig:device}
\end{figure}

A schematic figure of a molecule of DNA coupled to metalic electrodes is shown
in Figure  \ref{fig:device}. In summary, the source and drain electrodes are
taken into account
using self-energy functions \cite{Datta99}. The coherent
transport is studied by computing $G^r$, the retarded Green's function,
between source and drain, via a recursive lattice Green's function technique \cite{FG99}. These are the necessary ingredients which allow us to calculate the
density of states (DOS) and the transmission probability (T) used in our analysis of the localization
problem.

\subsection{The model Hamiltonians}

The model tight-binding Hamiltonian depicted in Figure  \ref{fig:device} 
belongs to a standard class of models, widely described in the
literature \cite{CunMRR07}, and can be written as: 

\begin{equation}
\label{eq:2channel}
\begin{array}{l}
H_{DNA}=
 \displaystyle\sum_{l=1}^{L}
 \Big[
  \sum_{\alpha=1}^{2}(\varepsilon_{l\alpha}| l,\alpha\rangle\langle l, \alpha |
- t^{x}_{l,\alpha}| l,\alpha\rangle\langle l+1,\alpha|)\\ 
  -t^{d}_{l}|l,1\rangle\langle l+1,2| - t^{d}_{l}|
l,2\rangle\langle l+1,1| - t^{y}_{l}| l,1\rangle\langle l,2|\\
  +\displaystyle\sum_{\sigma=\uparrow,\downarrow}(\varepsilon_{l}^{\sigma}|
l,\sigma\rangle\langle l,\sigma|-t_{l}^{\sigma}| l,\alpha(\sigma)\rangle\langle
l,\sigma |)\Big]
  +h.c..
\end{array}
\end{equation}

The different terms describe the on-site effective orbitals of the nucleotides,
the effective sugar and
phosphate orbitals and the hopping terms:
longitudinal along and transversal to the chains and the crossed (diagonal)
hopping between second nearest neighbors nucleotides \cite{ShiWHC12}. Here we follow 
the hamiltonian description given in a previous work \cite{CarpNR12}.

Hence, $t^{x}_{l,\alpha}$ is the hopping of base pair $l$ with first nearest neighbors along the strand
starting from 5' ($j=1$) and 3' ($j=2$) ends, $t^{d}_{l}$  denotes the diagonal
hopping and $t^{y}_{l}$ the
hopping perpendicular from 5' down to 3' at $l$. The sum over $\sigma$ in
equation(\ref{eq:2channel}) indicates the connection to the sugar-phosphate
backbone. In addition, $\alpha(\sigma)=1$ ($2$) for
$\sigma=\uparrow$ ($\downarrow$) and
$\epsilon_{l\alpha}$ and $\epsilon^{\sigma}_{l}$ denote the onsite energies on
the $2$ DNA strands and the top and bottom backbones.

The
onsite energies $\epsilon_{l\alpha}$ are taken to be the effective primary
ionization energies  of the base nucleotides, i.e. $\varepsilon_A=8.24$eV,
$\varepsilon_T=9.14$eV, $\varepsilon_C=8.87$eV and $\varepsilon_G=7.75$eV. In
this work, we consider the  backbone energy to be given as average of the
energies of the base nucleotides, i.e.
$\varepsilon_{l}^{\uparrow(\downarrow)}=8.5$eV for all $l$.
Both strands of DNA and the backbone are modelled explicitly and the different
diagonal overlaps of the larger purines (A,G) and the smaller pyrimidines (C,T)
are taken into account by suitable inter-strand
couplings \cite{WelSR09,RakVMR02}. Therefore, couplings are
$t^x_{l,\alpha}=0.35$eV between identical bases and $0.17$eV
between different bases; the diagonal inter-strand couplings are
$t^{d}_{l}=0.1$eV for
purine-to-purine, $0.01$eV for purine-pyrimidine and $0.001$eV for
pyrimidine-to-pyrimidine. The couplings to the backbone sites are
$t^{\sigma}_{l}=0.7$eV, and the hopping across a base pair is $t_{l}^{y}=0.005$eV. For previous
discussions leading to these choices of parameters as well as the influence of
the environment on the charge migration properties of the models, we refer the
reader to the existing literature \cite{Cha07,BerC07,KloRT05,CarpNR12}.

This model Hamiltonian is choosen for the reason that, on one hand it grasps the qualitative physics of charge
 transport in the molecule and, on the other hand it represents a quite general model taht can recover simpler ones 
 studied previously throughout the literature. 

Nevertheless, we emphasise that the choice of the tight binding parameters is
far from uniquely determined, being a rather controversial issue, and several
parameter sets have been proposed in the literature \cite{Roc03}.
\subsection{Green function techniques}

Having in mind the model Hamiltonian, localization and transport properties are obtained from the transmission probability between the electrodes: 

\begin{equation}
\label{eq:transmission}
\begin{array}{l}
        T(E)=Tr\left[ \Gamma_S G^r\Gamma_{D} (G^r)^\dag
\right],
\end{array}
\end{equation}
where $G^r$ is the retarted Green function of the system which can be found
from \cite{Datta99}
\begin{equation}
\label{eq:GR}
\begin{array}{l}
   G^r=[E-H_{DNA}-\Sigma_{S}-\Sigma_{D}]^{-1}.
\end{array}
\end{equation}

The self-energies $ \sum_{{S}({D})}=\tau_{{S}({D})}^\dag g_{{S}({D})}\tau_{{S}({D})}$  and the broadening function $\Gamma_{{S}({D})}=i\left(\Sigma_{{S}({D})}-\Sigma_{{S}({D})}^\dag\right)$ \cite{Datta99} are calculated as usual from
electrode Green's function $g$ (calculated using a recursive
technique \cite{LopLS85}) and the coupling $\tau$ between the DNA molecule and
electrode. The electrode-molecule coupling $\tau$ is determined by
the geometry of the chemical bond \cite{LarBD97}. We use 
couplings ranging \cite{PorBVD00} from  $\tau=0.35$eV to $\tau=1.5$eV. 

The diagonal elements of the $Im[G^r(i,j,E)]$ give us the local density of states (LDOS),
another important quantity for investigating the localization properties:
 \begin{equation}
\label{eq:dos}
\begin{array}{l}
   \rho(i,j,E)=-\frac{1}{2\pi}Im[G^r(i,j,E+i\gamma)].
\end{array}
\end{equation}

Considering the present open systems and the limit $\gamma\rightarrow0$, 
one may establish a relation between the LDOS and the electronic probability density:  
$\rho(i,j,E)=\Sigma_\alpha \delta(E-\varepsilon_\alpha) |\psi(i,j)|^2$, where $\psi(i,j)$ is
 defined as the eigenstate with eigenenergy $\varepsilon_\alpha$, of the effective hamiltonian $H=H_{DNA}+\sum_{S}+\sum_{D}$.

\section{Localization length and participation number}
\label{sec:Theory}

A further and fundamental step undertaken here is the analysis of the
localization of the
electronic
states. The concept of localization has been originally defined for the
thermodynamic limit, but the advent of mesoscopic systems brought new aspects to the
understanding of the localization properties. Considering an increasing degree
of disorder in a mesoscopic system, the transport can be tuned from ballistic
down to the localized regime, with a diffusive transport window between both
limits. The
LL is a relevant length scale defining these regimes \cite{GuhMW98}; in particular, one has a localized regime when L, the length of
the system, is much larger than LL: $L\gg LL$. The
diffusive and ballistic regimes are partially characterized by $L\ll LL$.
Furthermore, the
transition from localized to diffusive transport is rather a wide crossover in
the range $L\approx LL$. Hence, the degree of localization for mesoscopic
systems has to be carefully investigated. A possible approach is
to calculate the LL in the thermodynamic limit and then compare
the result with the characteristic length of the mesoscopic system. Such
approach, although able to define if the system is far away from the crossover
around $L\approx LL$, either in the
localized or diffusive regimes, is inconclusive concerning the transition
between both regimes. Therefore we analyze here the evolution of different
definitions of the degree of localization as a function of the system size in
order to identify the transition from localized to diffusive regimes in
different DNA-like heuristic models.

The quantities described in the previous section, T(E) and $\rho(i,j,E)$, are
relevant ones in defining the localization of a electronic state in the
disordered central DNA-like ladder model.

On one hand, LL is computed from the
exponential decrease in the transmission probability \cite{Kis91,JohK83}:

\begin{equation}
\label{eq:LL}
\begin{array}{rl}
  
LL^{-1}(E)=\displaystyle-\lim_{L\to\infty}\frac{1}{2L}<\ln T(E)>,
\end{array}
\end{equation}

where $<\cdot\cdot\cdot>$ means an average over several hundred different
disordered chain
configurations.  This is done to avoid spurious resonance effects due to a
particular
configuration. Here $L$ is the length of the system given in number of bases
pairs. 
Hence, for the sake of completeness in the forthcoming discussion, the
total number of effective sites in the system is 4L.

Alternatively, another way to define the localization degree of an electronic
state is obtained directly from the wave function, namely, the participation
ratio,  defined initially by \cite{EdwT72,HofS93} for finite systems.
Here we define a participation ratio with open boundaries $PR_{OB}$, 
counting the contributions at each site to the density of states at a given
energy:

\begin{equation}
\label{eq:PR}
\begin{array}{l}
  
PR_{OB}(E)=\displaystyle\frac{1}{4L\displaystyle\sum_{i=1}^{L}\sum_{j=1}^4
|\rho_{ij}|^2}.
\end{array}
\end{equation}

For
localized states, $PR_{OB}\rightarrow0$ in the limit $L\rightarrow \infty$,
while
truly delocalized states will lead to  values up to $PR_{OB}\rightarrow1$. A
quantity related
to the $PR_{OB}$ is the Particpation number (PN), here, PN=$4L$ $\cdot$
$PR_{OB}$. While $PR_{OB}$ is simply a
fraction, PN would be a measure of the actual number of sites having
appreciable wave-function amplitudes at a given energy. 

A final step is the investigation of the effects in changing
the coupling of the DNA-like device to the leads. Such investigation is, on one hand, a simple
heuristic approach to 
the effects of the contacts to the leads on the actual transport properties \cite{Datta99}. 
On the other hand, combining the evolution of different
definitions of the degree of localization, as a function of the system 
size, with the sensitivity to boundary conditions (by changing the contacts), 
further way to define electronic localization \cite{EdwT72}, will lead to a
comprehensive picture of the electronic state extension properties in mesoscopic
length scales.

\section{Results and Discussion}
\label{sec:results}

\begin{figure}[t!]
\begin{center}
\includegraphics[width=0.55\columnwidth]{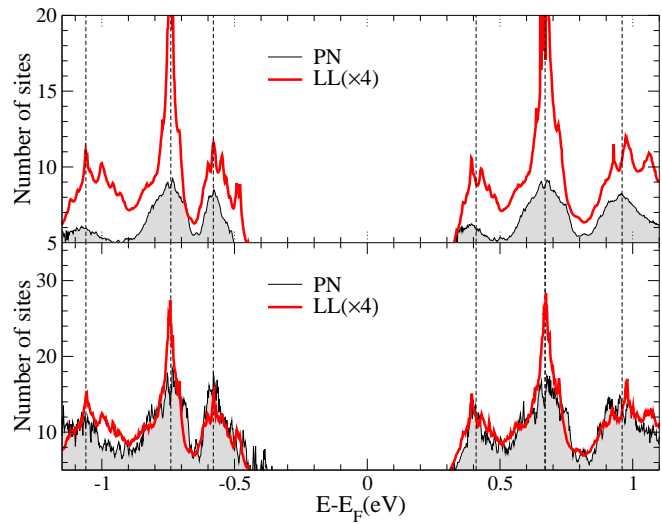}
\end{center}
\vspace{-0.2cm}
\caption{(color online): Localization length (LL) and participation number (PN) as a function of energy for DNA-like ladders with backbones, $L=30$ (top)
bps and $L=300$ bps(bottom) long. The vertical lines are guidelines identifying robust structures associated to resonances in the transmission pobabilities.}
\label{LC_NP_MN_300}
\end{figure}

A general picture of the question raised in the present work is given by a first
comparison between LL and PN, as a function of energy, shown in 
figure  \ref{LC_NP_MN_300} considering two different disordered systems
lengths: 30 base pairs (bps) and 300 bps (corresponding to 10.4 nm and 104 nm,
respectively). The depicted results here and in the following figures, with the
exception of figure  \ref{Localdensity}, are averages over hundreds of disorder configurations. A
first main feature is the localization gap, resembling the semiconducting gap
introduced by adding backbones to a ladder model, as reported in previous works \cite{KloRT05, MacR06}, and absent in simple double chain models \cite{CCPD02}.
Furthermore, considering the 300 bps long case, all the states are localized,
showing a LL an order of magnitude smaller than the system
length.

A central issue refers to the differences between the LL, obtained from the
transmission probability by means of equation(\ref{eq:LL}), and the PN, obtained as described
above, starting from the participation ratio, equation(\ref{eq:PR}). Having
in mind that PN is a well defined quantity irrespective to the system size,
while LL may be ill defined below the thermodynamic limit, it is worth noting that a
gross qualitative agreement between both quantities may be inferred even at the
30 bps case at the upper panel of figure  \ref{LC_NP_MN_300}. However, from a
quantitative point of view, the disagreements are significant. A much better
agreement, both qualitative and quantitative, is obtained for the 300 bps long
system. 

Two important aspects should be kept in mind for the discussion in the
following. (i) The scaling between LL and PN, considering that PN embraces not
only the system length, but also the number of chains, not all of them coupled
to the contacts and taking part in the transmission. (ii) The quantitative
agreement is not uniformly achieved along the entire energy range, particularly
at the two anomalously large LL peaks, introduced by the backbones, but also
present in other related systems like
G4-DNA \cite{GuoX09}.

A deeper understanding on these aspects and the related definitions of
localization is obtained by inspecting of the LDOS for single disorder
configurations at chosen values of energy.
In figure  \ref{Localdensity} one can observe the LDOS
for single disorder realizations at three energies, 
E =-1.05 eV, E=-0.74 eV and E =-0.58 eV, which correspond to the highest values
of the three peaks in the LL for the valence band in the lower
panel of figure  \ref{LC_NP_MN_300}, i.e., DNA strands 300 bps long. The panels show the LDOS along the system, discriminating the contribution of each of
the 4 chains: two chains for the backbones and two constituted by the
nucleotides. 

\begin{figure}[t!]
\begin{center}
\includegraphics[width=0.55\columnwidth]{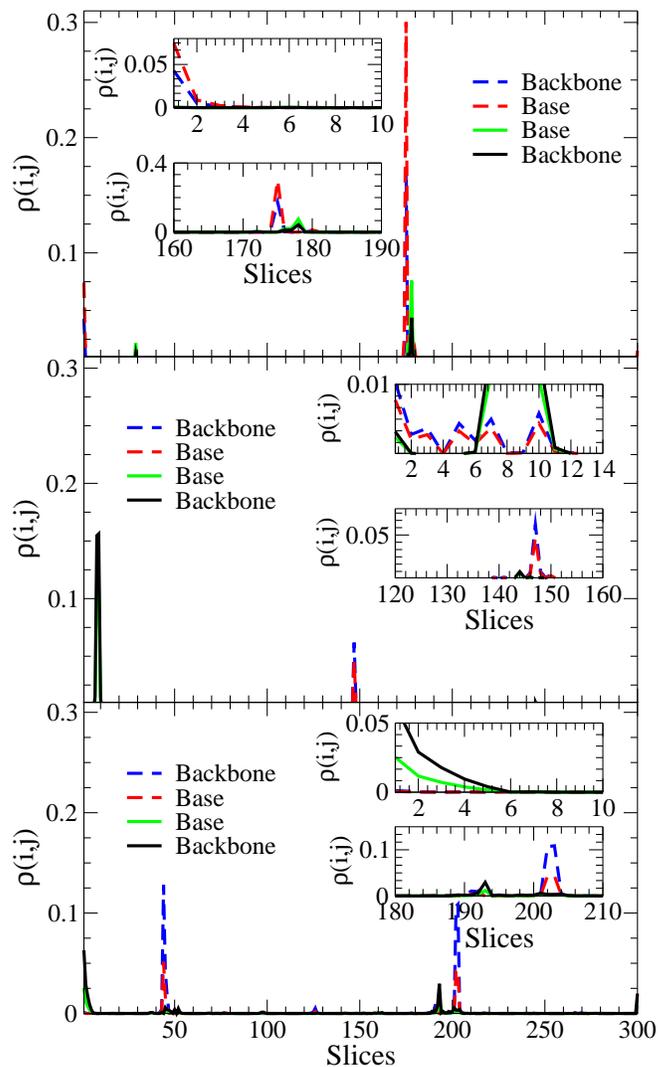}
\end{center}
\vspace{-0.2cm}
\caption{(color online): Local density of states $\rho(i,j)$ along all the base pair slices of a 300 bps double chain at the energies
E =-1.05 eV (top), E=-0.74 eV (middle) and E
=-0.58 eV (bottom), which correspond to the longest LL in the spectrum shown in bottom panel of figure  \ref{LC_NP_MN_300}. 
The contributions to $\rho(i,j)$ 
from the four
chains of the system are discriminated:  
external backbone chains (blue and black) and the internal base pair like ones (red and
green). The insets are zooms at the begining of the chains near the source contacts (top insets) and at a chosen resonance region (bottom insets).}
\label{Localdensity}
\end{figure}

All three cases show LDOS with resonances which may appear at any position,
sometimes deep inside the system, far from the contacts. From the insets one apprehends that such
resonances are indeed very sharp, involving a small number of effective sites.
One also observes overall
exponentially decaying states from the contacts into the DNA-like double chain
with attached backbones, as can be seen from the insets showing the first few
base pairs of each case. The tuning of these resonances may enhance dramatically
the transmission probability at certain energies for a particular disorder
configuration, but, on average, such effects should be washed out. Nonetheless,
the inspection of single disorder configurations reveals that the presence of
resonances are quite common, meaning that the incident electronic wave function
may couple to states inside the system localized at distances much larger than
the scale given by LL. Some configurations, not shown here, may show up to several peaks in the LDOS without a well defined envelope. These findings are qualitatively
different from previous ones for bare base pairs ladders (absence of backbones) \cite{CarP08}, 
for which much longer LLs are observed with
resonant states spread out over a large number of sites. In spite of these
differences, the presence of resonances  (in the present case, the resonances
are sharpened by the presence of the backbones), raises the question of defining
a new length scale,
given by $L_{wf}>LL$, which may elucidate the effect of resonance coupling on
the effective localization of states. 

\begin{figure}[Ht]
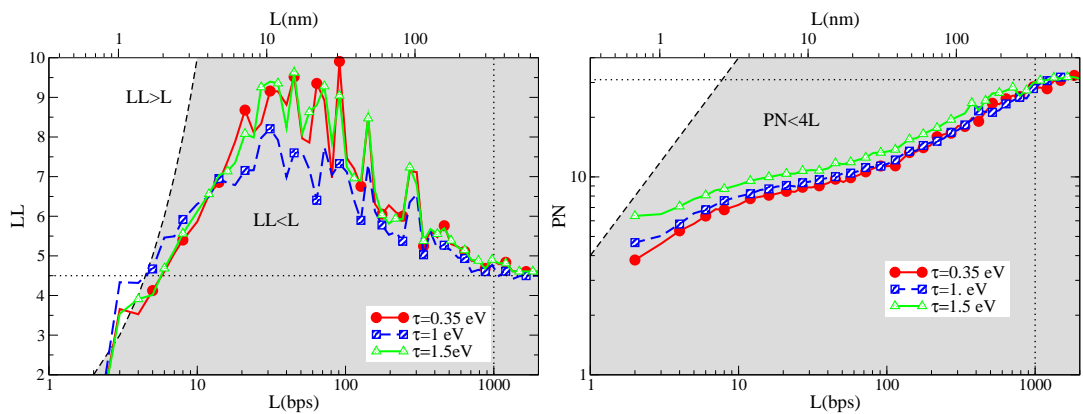

\begin{center}
\includegraphics[width=0.45\columnwidth]{Fig_4a.eps}
\includegraphics[width=0.45\columnwidth]{Fig_4b.eps}
\end{center}
\vspace{-0.2cm}
\caption{(color online):  Localization length (LL)(left panel) and participation number (PN) (right panel) as
a function of double-chain length for different double-chain to contact coupling strengths, $\tau$, at E=-0.74 eV, corresponding to the highest LL in the 
valence band in figure  \ref{LC_NP_MN_300}.
The chain lengths are given in units of bps, as well as in nanometers, and the coupling strengths considered are: 
$\tau=0.35$ eV (red) $\tau=1$ eV (blue) and $\tau=1.5$ eV (green).The shaded areas ($LL < L$ and $PN < L$) 
correspond to the localized regime, while effective delocalization occur at the white areas.}
\label{fig:LLvsm}
\end{figure}

The closer quantitative agreement between PN and LL for longer systems suggests
that the effect of such resonances become progressively less important with
increasing system length. Therefore, the transition from short to longer
systems, concerning the degree of localization has to be carefully investigated,
as illustrated in figure  \ref{fig:LLvsm}, where LL (left panel) and PN (right
panel), as a function of the
system length at the energy of the highest LL peak at the valence band in figure  \ref{LC_NP_MN_300}, are shown. The different curves are for different intensities of the coupling
to the contacts, introducing the sensitivity to the boundary conditions and, therefore, a size scaling analysis.  Several
aspects can be appreciated in these results. First of all, differently than for other
related heuristic models \cite{CarP08} there is no window of effective
delocalization, as can be observe that the numerical results for the degree of
localization never reach the white region where the LL would be
larger than the system size. On the other hand, the results for LL and PN become
length independent at lengths beyond 1000 bps. Two other aspects may call the
attention here. First, the LL initially increases with the system length,
reaching a maximum and saturating at a lower value at the thermodynamic limit,
which for the present case is reached at 300 nm (upper scale of the left panel
in figure  \ref{fig:LLvsm}). Second, the PN shows a steady increase with the system length, but
never reaching the necessary steepness (dashed line at the right panel)
representing effective delocalization.  Furthermore, it is also noticeable that
the sensitivity to the contact coupling is relevant up to lengths of hundreds of
bps, the results becoming insensitive to the coupling intensity for lengths
beyond 1000 bps.

From the point of view of actual transport measurements in DNA systems, the coupling to
the contacts sensitivity of the LL, a quantity directly related to the
transmission probability, is expected to be of relevance within a
manifold of other environmental variables. This apparently obvious statement
becomes more involved when one also observes the strong dependence of the PN on
the contact coupling, since variations in the PN suggest modifications in the
LDOS distribution at the given energy, hence the charge distribution in the
system.

The results shown in figure  \ref{fig:LLvsm} suggest that three ways of
defining the degree of localization reach insensitivity in respect to the system
size at roughly the same system size, defining a the thermodynamic
limit in this context of 300 nm, i.e., within a mesoscopic lenght scale.

\begin{figure}[Ht]
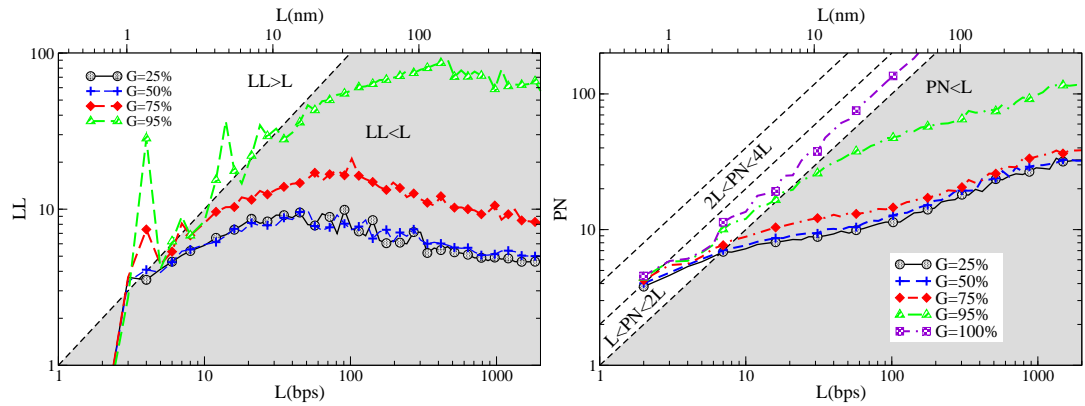

\begin{center}
\includegraphics[width=0.45\columnwidth]{Fig_5a.eps}
\includegraphics[width=0.45\columnwidth]{Fig_5b.eps}
\end{center}
\vspace{-0.2cm}
\caption{(color online):  
Localization length (LL)(left panel) and participation number (PN) (right panel) as
a function of double-chain length, at E=-0,74 eV, for different concentrations of G nucleotides: 25\% (black), 50\% (blue), 75\% (red), 95\% (green) and 100\% (violet), 
ranging hence from strong disorder ( 25\% and 50\% ) to almost ordered (95\%) and completely ordered (100 \%). The shaded areas (LL < L and PN < L) 
correspond to 
the localized regime, while effective delocalization occur at the white areas}
\label{fig:LLvsm_pc}
\end{figure}

A transition from localized to diffusive transport regime would imply in a
crossing of LL from the grey to the white area, or a PN showing the same
steepness of the dashed border line, in figure  \ref{fig:LLvsm} for a certain length range.
Evidences of such behavior appear in other related models \cite{CarP08}
characterized by quite longer LLs, but are clearly absent here. 
Nevertheless, the disorder considered here is the completely random choice of nucleotides,
solely constrained by base pairing, hence there is on average a 25\% 
concentration of each of the bases, A,T,C and G. By increasing the concentration
of G, for instance, the degree of disorder may be reduced, reaching eventually
an ordered system at the limit of 100\% concentration of G. This dependence to
the degree of disorder is shown in
figure  \ref{fig:LLvsm_pc} where LL and PN are depicted as a function of
size, 
for several concentration of G-like nucleotides at a central chains. Either PN
and LL increase with the concentration of G, i.e., diminishing the degree of
disorder. Indeed, for the low disorder case of G corresponding to 95\% of the
sites, a diffusive regime may be observed for the present model Hamiltonian. A
completely ordered system would imply in transmission probabilities close to the
unity, leading to numerically instabilities in defining the LL by means of equation \ref{eq:LL}
. However, the PN can be easily defined for the ordered system and this case is
included in figure  \ref{fig:LLvsm_pc} and it is worth noting that it follows
the expected steepness but in the range $L<PN<2L$, indicating that the LDOS is
not evenly distributed among the 4L sites of the system.

\section{Conclusions}
\label{sec:conclusions}
The present work addresses the question of characterizing the degree of
localization in finite disordered systems, considering the case of a more
realistic,
although heuristic, model for finite DNA-like systems. The degree of
localization of electronic states in small systems is relevant in the sense that
it could show transport properties of a localized regime in the thermodynamic
limit, while a shorter chain could fall within a diffusive regime. This is
indeed the case for other models and parametrizations within the 
same class for which overall longer LL appear \cite{CarP08}. In the present
case, a crossover to the $LL>L$ range is only achieved by reducing the degree
of disorder. Underneath
this potentially interesting possibility of mesoscopic systems, represented here
by DNA-like chains, a more fundamental problem has been addressed, namely the
behavior of various definitions of localization at lengths below the
thermodynamic limit. The present results elucidate the discrepancies among the different
definitions of localization at small system sizes, below the thermodynamic limit. These discrepancies can be
attributed to the resonances in the system, that play 
a decreasing role with increasing system length. For the present case, from the
point of view of the localization properties, the thermodynamic 
limit present a threshold in the mesoscopic range, namely at lengths
around a few hundreds of nanometers. Finally, having in mind the connection 
of such heuristic models to real DNA systems, the present results give a simple picture
 to the great variety among the first experimental results on DNA 
transport properties \cite{EndCS04} realized in the length scales studied here:
rather small changes in the degree of disorder, contact strength and electronic
structure parameters may considerably modify the transport properties of a modeled
device.

\ack


The authors are grateful to Rudolf A. Roemer for a critical reading of the manuscript. C.J.P.\ acknowledges financial support from FAPESP (Brazil) and  P.A.S.\ acknowledges partial support
from CNPq (Brazil).

\section*{References}

\bibliographystyle{bibliography/unsrt}
\bibliography{bibliography/bibliograph}

\end{document}